\documentclass[conference,onecolumn]{IEEEtran}
\usepackage{amssymb}
\usepackage{amsmath}
\usepackage{lineno}
 \usepackage{float}
\usepackage[ruled,vlined]{algorithm2e}
\usepackage{cite}
\usepackage{tipa} 
   \renewcommand{\baselinestretch}{1}\normalsize
 \ifCLASSINFOpdf
   \usepackage{graphicx}
 \else
    \usepackage[dvips]{graphicx}
   \graphicspath{{New Volume/wRITINGS/recokon/}}
 \fi

\begin{document}

Note: This paper is a preprint of a paper accepted by IET Wireless Sensor Systems and is subject to Institution of Engineering and Technology Copyright. The copy of record is available at IET Digital Library

\title{Dead Reckoning Localization Technique for Mobile Wireless Sensor Networks}

\author{ \footnotesize \IEEEauthorblockN{Haroon Rashid}
\IEEEauthorblockA{Department of Computer Science and Engineering\\
National Institute of Technology\\
Rourkela, Odisha -- 769008\\
Email: haroon.it@gmail.com
}
\and
\IEEEauthorblockN{Ashok Kumar Turuk}
\IEEEauthorblockA{Department of Computer Science and Engineering\\
National Institute of Technology\\
Rourkela, Odisha -- 769008\\
Email: akturuk@nitrkl.ac.in}
}
\maketitle
 \renewcommand{\baselinestretch}{1.5}\normalsize
\thispagestyle{plain}
\pagestyle{plain}
\begin{abstract}
\noindent
Localization in wireless sensor networks (WSNs) not only provides a node with its geographical location but also  
a basic requirement for other applications such as geographical routing. Although a rich literature is available for localization in static WSN, not
enough work is done for mobile WSNs, owing to the complexity due to node mobility. Most of the existing
 techniques for localization in mobile WSNs uses Monte-Carlo localization (MCL), which is not only time-consuming but also memory intensive. They, 
consider either the unknown nodes or anchor nodes to be static. In this paper, we propose a technique called Dead Reckoning Localization for mobile
WSNs (DRLMSN). In the proposed technique all nodes (unknown nodes as well as anchor nodes) are mobile. Localization in DRLMSN is done at discrete time 
intervals  called checkpoints. Unknown nodes are localized for the first time using three anchor nodes. For their subsequent localizations, only two anchor nodes are used. 
 The proposed technique estimates two possible locations of a node Using B\'ezout's theorem. A dead reckoning approach is used to select one of the two estimated locations.
 We have evaluated DRLMSN through simulation using Castalia simulator, and is compared with a similar technique called RSS-MCL proposed by Wang and Zhu \cite{comp}.
\end{abstract}
\begin{IEEEkeywords}
 Mobile wireless sensor networks, Distributed localization, RSSI, Range based localization, Dead reckoning, Trilateration.
\end{IEEEkeywords}
\IEEEpeerreviewmaketitle
\section{Introduction}
\noindent
Localization is the process of providing each node with its geographical location. It is an important
 research issue
 in wireless sensor networks. This is because, 
each sensor node stamps the sensed data with its geographical location prior to transmission.
The sink, distinguishes the received data based on the spatio-temporal characteristics, in which the location information is  
 important and has a unique characteristic. Therefore, the transmitting node must be aware of its geographical position.

 Localization not only provides the 
geographical position of a sensor node but also fills the pre-requisites for geographical routing, load balancing, spatial querying, data dissemination,
rescue operations, and target tracking etc \cite{newapp1,newapp2,patwari,iet1,rescue}.
Location based routing can save significant amount of energy by eliminating the need for route discovery \cite{mobile}. It can also improve the caching behaviour of
 applications where the requests are location dependent.

Mobility of sensor nodes increases the applicability of wireless sensor networks (WSNs). The mobility of a sensor node with respect to environment can be of two
 types: \textit{(i)} static, and \textit{(ii)} dynamic.
 In static, a sensor node is only data driven, i.e., to sense the environment and report to the base 
station (BS), while in dynamic a sensor node is not only data
driven, but also serves as an actuator.
In both the cases, position of nodes changes oftenly.
As a result, a node in mobile WSNs is localized more than once, compared to static WSNs - where a node is
localized only once at the initialization of network. The continuous localization of nodes with mobility results in: \textit{(i)} faster battery deletion and hence,
reduces the lifetime of sensor nodes, and \textit{(ii)} increases the communication cost. On the contrary, mobility improves: \textit{(i)} coverage of WSN - uncovered
locations at one instant can be covered at some other instant of time, \textit{(ii)} enhances the security - intruders can be detected easily as compared
to the static WSNs, and \textit{(iii)} increases connectivity - mobility increases the neighbours of a node \cite{book1}.

The geographical position of a sensor node is determined either with the help of global positioning system (GPS) equipment or estimated from the location of neighbouring nodes.
A network with all GPS enabled nodes is not a universal solution. This is because besides increasing the cost, size, and power consumption; it 
fails to localize in indoors, and dense forests. An alternative solution is to equip only a minimal number of nodes called \textit{anchor nodes/seeds}
 with GPS, and the remaining nodes estimate their location using the location information of anchor nodes. 
  A mobile WSN can be in one of the following scenarios \cite{mobile,onebeacon2,onebeacon3,onebeacon4,onebeacon1,mazzimi,centroidm,jangping,wanggroup,jamalkak,
  savic,springerist,thesis,springer2}:
\begin{enumerate}
 \item \textit{Normal nodes are static, and seeds are moving}: In this scenario, mobile anchor 
nodes ($\geq 1$) continuously broadcast their location.
As soon as a static node receives three or more beacons, it localizes itself. Accuracy and localization time depends
mostly on the trajectory followed by the seeds.
\item  \textit{Normal nodes are moving, and seeds are static}: In this scenario, each normal node is expected
to receive the beacons at the same instant of time. Otherwise, it results into inaccurate estimated location. With time span the previous
estimated location becomes obsolete. As a result, nodes localize repeatedly at fixed intervals with newly received seed locations. One
of the appropriate example for this scenario is battlefield, where normal nodes are attached to military personnel and seeds are fixed a priori as landmarks
within the battlefield. This not only helps in detecting the current position but also in providing feedback from a particular area of 
battlefield.
\item \textit{Both the normal nodes, and seeds are moving}: This scenario is the most versatile, and complex among all the three. In this, the
topology of the network changes very often. It is difficult for a normal node to get fine grained location. Therefore, the localization
error is comparatively higher than the previous two scenarios.      
\end{enumerate}
   In this paper, we propose a localization technique for the  scenario,  where both the normal nodes and seeds are moving. We have considered this scenario because: 
 \textit{(i)} a little emphasis is given, owing to its complexity, and \textit{(ii)} to the best of our
knowledge whatever little localization techniques exist for this scenario are range free based.  
 The proposed technique is called Dead Reckoning Localization Technique (DRLMSN). Due to node mobility the position of nodes changes frequently. Localizing the
 nodes at every instant will drain their battery power at a faster rate. Therefore, in DRLMSN nodes are localized at discrete time intervals called checkpoints.
 Nodes are localized for the first time using trilateration mechanism. In their subsequent localizations, only two anchor nodes are used. B\'ezout's theorem \cite{bezo} is used
 to estimate the positions of a node, and a dead reckoning technique is used to select the correct position. DRLMSN is compared with RSS-MCL \cite{comp}.
 
The rest of the paper is organized as follows: Section \ref{lit} provides an overview of various localization techniques. Proposed
localization algorithm is described in Section \ref{model}. Simulation and results are presented in Section \ref{eval}, and conclusions
are drawn in Section \ref{con}.
\section{Related work} \label{lit}
\noindent
Localization algorithms for WSNs can be broadly categorized into two types: \textit{Range based}, and \textit{Range free}. Range based localization algorithms uses 
range (distance or angle) information from the beacon node to estimate a sensor node's location \cite{types}. They need at least three beacon nodes to estimate
the position of a node. Several ranging techniques exist to estimate the distance of a node to three or more beacon nodes. Based on this information,
 location of a node is determined. A few representative of range based localization algorithms include: Received signal strength indicator (RSSI) \cite{rssi2}, 
Angle-of-arrival (AoA) \cite{tdoa}, Time-of-arrival (ToA) \cite{toa}, Time difference of arrival (TDoA) \cite{tdoa}. These schemes often need extra hardware such as
antennas and speakers, and their accuracy is affected by multi-path fading and shadowing.

Range-free localization algorithms use only connectivity information between unknown node and landmarks - which obtain their location information using
 GPS or through an artificially deployed information. Some of the range-free localization algorithms include: Centroid \cite{triangle}, Appropriate point
in triangle (APIT) \cite{apit}, and DV-HOP \cite{dvhop}. Centroid counts the number of beacon signals it has received from the pre-positioned beacon nodes and 
achieves
 localization by obtaining the centroid of received beacon generators. DV-HOP uses location of beacon nodes, beacon hop count, and the average distance
per hop for localization. APIT algorithm requires a relatively higher ratio of beacons to nodes and longer range beacons for localization. They are
 susceptible to erroneous reading of RSSI. He \textit{et al.} \cite{apitsupport} showed that APIT algorithm requires lesser computation than other beacon based algorithms and
performs well when the ratio of beacon to node is higher. A brief review of different localization algorithms proposed in the literature for mobile sensor networks
 is presented below:

In \cite{tilak}, authors proposed two classes of localization approaches for mobile WSNs. They are: \textit{(i)} Adaptive, and \textit{(ii)} Predictive.  
 Adaptive localizations dynamically adjust their localization period based on the recent observed
motion of the sensor, obtained from examining the previous locations. This approach allows the sensors to reduce their localization frequency when the sensor is slow, or
increase when it is fast. In the predictive  approach, sensors estimate their motion pattern and project their future motion. If the prediction is
accurate, which occurs when nodes are moving predictably, location estimation may be generated.
However, their main focus was on how often the localization should be done, and not on the localization process itself.

Bergamo and Mazzimi \cite{mazzimi} proposed a range based algorithm for localizing mobile WSNs. They used fixed beacons placed at two corners on the same side of 
a rectangular space. Mobile sensor uses beacon signals to compute their relative position. Sensors estimate their power level from the
 received beacon and compute their position by triangulation method. They also studied the effect of mobility and fading on
 location accuracy. Placing beacons at fixed position limits the localization area. This is because the quality of RSSI decreases with distance due to various
 factors. This results in inaccurate distance measurement.
 
Hu and Evans \cite{mobile} proposed a range free technique based on Monte-Carlo Localization (MCL). This technique is used for localization
of robots in a predefined map. It works in two steps: First, the possible locations of an unknown node is represented as a set of weighted samples. In the next 
stage, invalid samples are filtered out by incorporating  the newly observed samples of seed nodes. Once enough samples are obtained, an unknown node estimate its 
position by taking the weighted average of the samples. In this technique, the sample generation is computationally intensive and iterative in nature.
 This also needs a higher seed density.

Aline \textit{et al.} \cite{aline} proposed a scheme to reduce the sample space generated in \cite{mobile}. They named it as Monte-Carlo Boxed (MCB) scheme. 
Sample generation is restricted within the bounding box, which is built using 1-hop and 2-hop neighbouring anchor nodes. The neighbouring anchor information 
is also used in the filtering phase. The number of iterations to construct the sample space is reduced. However, the localization error in MCB is not
reduced, if the number of valid samples is same as that in MCL.

Using MCL technique, Rudafshani and Datta \cite{rudaf} proposed two algorithms called MSL and MSL*. Static
sensor nodes are localized in  MSL* too. 
 It uses sample set of all 1-hop and 2-hop neighbour
 of normal nodes, and anchor nodes. This resulted in better position estimation with increased memory requirement and communication cost. 
 In MSL, a node weight its samples using the estimated position of common neighbour nodes.
MSL* outperforms MSL in most scenarios, but incurs higher communication cost. MSL outperforms MSL* when there is significant irregularity in the radio range.
Accuracy of common neighbouring nodes are determined by their closeness value. Closeness value of a node P with N samples is computed as:
\begin{center}${Closeness_P} =\dfrac{\sum_{i=1}^{N} W_i \sqrt{(x_i - x)^2 + (y_i - y)^2}} {N}$\end{center} 
where ($x,y$) is P's estimated position and ($x_i,y_i$) is P's $i^{th}$ sample with weight $W_i$. Both MSL and MSL* needs higher anchor density and node density.
Also, when maximum velocity is large, performance of both MSL and MSL* reduces to a greater extent. Furthermore, the
size of bounding box for sample generation is reduced in \cite{shigeng}, using the negative constraint of 2-hop  neighbouring anchor nodes. This 
reduces the computational cost for obtaining samples, and a higher location accuracy is achieved under higher density of common nodes. 

Wang and Zhu \textit{et al.} \cite{comp} proposed the RSS based MCL scheme which sequentially estimates the location of mobile nodes. First, it uses a set of samples with related weights to represent
the posterior distribution of node's location. Next, it estimate the node's location recursively from the RSS measurement within a discrete state-space localization system. 
Accuracy of this scheme depends on the number of samples used and the log normal statistical model of RSS measurements. It improves the localization accuracy  
using both the RSSI and MCL techniques. However, it is time intensive to the sampling procedure. Also, the frequent sampling at regular intervals consume more energy of 
the nodes.

Comparison of different localization techniques is shown in Table \ref{tab:xyz}.
\begin{table}[h]
\centering
\small
\begin{tabular}{|p{2.8cm}|p{1.7cm}|p{2.4cm}|p{4.5cm}|p{4.4cm}|}
\hline
{\bf Localization Technique }&{\bf Type }&{\bf  Mechanism} & {\bf{ \centering Mobility Model}} & {\bf Comments}\\ \hline \hline 
  Tilak et al. \cite{tilak} & Range Free & Triangulation &  Random Waypoint Model (RWP), Gaussian Markovian Model &
  Emphasized on localization frequency rather than localization accuracy.\\ \hline
  Bergamo and Mazzimi \cite{mazzimi}& Range Based  & Triangulation & RWP & Limits the localization area, consider static beacon nodes.\\ \hline
  Hu and Evans \cite{mobile} & Range Free& Sequential MCL & RWP, Reference Point Group Model \cite{refmodel} & Slow in convergence, slower sampling technique.\\ \hline
  Baggio and Langendoen \cite{aline} & Range Free& MCL Boxed & Modified RWP with pause time = 0 \& minimum node speed = 0.1 m/s. &Lesser the sample space, more the localization error.\\ \hline
  Rudafshani and Datta \cite{rudaf} & Range Free& Particle filtering  approach of MCL & Modified RWP with pause time = 0 second. & Computationally intensive, higher communication cost, requires
  higher anchor \& node density\\ \hline
  Wang and Zhu \cite{comp} & Range Based & Sequential MCL & RWP & Accuracy depends on the sampling quality, and the RSSI model.\\ \hline
\end{tabular}
\caption{Comparison of different localization techniques for Mobile WSN.}
\label{tab:xyz} 
\end{table}
 \section{Dead Reckoning Localization Technique}\label{model}
 \noindent
In this section, we propose a range based, distributed localization algorithm for mobile WSNs. The proposed technique is called Dead Reckoning Localization 
Technique (DRLMSN). Nodes in DRLMSN are classified into the following three types: 
(\textit{i}) \textit{Anchor node} ($\mathcal{A}$): A node which can locate its own position, and is usually equipped with GPS,
(\textit{ii}) \textit{Normal/unknown node} ($\mathcal{U}$): Nodes which are unaware of their location, and uses localization algorithm to determine their position, and
(\textit{iii}) \textit{Settled node} ($\mathcal{S}$): These are normal nodes that have obtained their location information through a localization technique. They
serve as an anchor node for the remaining unknown nodes. 
 
To localize normal mobile nodes accurately with the help of mobile anchor nodes is a difficult task. This is because the transmitter as well
 as the receiver changes their position at every time instant. Therefore, to localize, a normal node must receive \textit{beacons} from all the neighbouring nodes
 at the same time instant. \textit{Beacon} are frequently advertised from the anchor/settled nodes. This advertisement contains the anchor/settled \textit{node's identiy}, 
and \textit{location}. Continuous localization of mobile nodes will drains their battery power at a faster rate. As a result, the lifetime of sensor nodes as well as the
 sensor network is reduced.
 
Sensor nodes in DRLMSN are localized during a time interval called  \textit{checkpoint}.
 There are two localization phases in DRLMSN. 
 First phase is called \textit{Initialization} phase. In this phase, a node is localized using trilateration mechanism. A node remains in the initialization phase 
 until it localizes using trilateration mechanism. The subsequent localization phase is called \textit{Sequent} phase. In this phase, a node localizes itself using only two anchor nodes.
B\'ezout's theorem \cite{bezo} is used to estimate locations of a node. A \textit{dead reckoning} approach is used to identify their correct estimated position. 
Once a node is localized in either of the above two phases, it act as a settled node and broadcast beacons during the checkpoint.
Initialization and Sequent phases are explained below.

\subsection{Initialization Phase:} \label{subsec1}
\noindent
At the beginning of a \textit{checkpoint}, each anchor node broadcasts a beacon. A normal node localizes itself for the first time 
during the checkpoint using three anchor nodes. As soon as, a node localizes, it broadcasts 
a beacon during the same checkpoint. This results in the localization of one/two beacon deficit nodes. This process continues until the end of the checkpoint.

At the end of the checkpoint, some nodes may fail to localize. The possible reasons for localization failure and the corresponding actions to be taken are: 
\textit{i)  A normal node receives only one (or two) beacon}. In this case, normal node deletes the received beacons and moves on. In the subsequent
 checkpoint it attempts to localize using three beacons.
\textit{ii) A normal node receives no beacon}. In this case, a node moves on and attempts to localize itself in the next checkpoint using three beacons.   
\subsection{Sequent Phase:} \label{subsec2}
\noindent
A node goes to the sequent phase only after itself localizes using trilateration mechanism.
In this phase, each normal node localizes with only two nearest location aware nodes (\textit{anchor / settled node}). A normal node that receives two beacons,
can estimate its two positions using B\'ezout's theorem.  According to  B\'ezout's theorem \textit{``The intersection of a variety of
degree $m$ with a variety of degree $n$ in complex projective space is either a common component or it has $mn$ points when the intersection points are counted 
with the appropriate multiplicity''}. Positions estimation of a node using  B\'ezout's theorem is explained below.

 Let $(x,y)$ be the position of an unknown node, and $(a_1,b_1)$, $(a_2,b_2)$ be the position of two of its neighbouring anchor nodes. Also, let
the distance between an unknown node, and the respective anchor nodes be  $d_1$ and $d_2$ respectively. Then,
\begin{align}
 (x-a_1)^2+(y-b_1)^2 = {d_1}^2 \label{eq1}\\
(x-a_2)^2+(y-b_2)^2 = {d_2}^2 \label{eq2}
\end{align}
Re-arranging (\ref{eq1}) and (\ref{eq2}), we obtain:
\begin{align}
x^2+y^2 =  {d_1}^2-{a_1}^2-{b_1}^2+2a_1x + 2b_1y \label{eq3}\\
x^2+y^2 =  {d_2}^2-{a_2}^2-{b_2}^2+2a_2x + 2b_2y \label{eq4}
\end{align}
Comparing (\ref{eq3}) and (\ref{eq4}), we have
\begin{align}
{d_1}^2-{a_1}^2-{b_1}^2+2a_1x + 2b_1y =  {d_2}^2-{a_2}^2-{b_2}^2+2a_2x + 2b_2y \label{eq5} \\
2(a_1-a_2)x = ({d_2}^2-{a_2}^2-{b_2}^2-{d_1}^2+{a_1}^2+{b_1}^2)+2(b_2-b_1)y \label{eq6}
\end{align}
Let $z_0= {d_2}^2-{a_2}^2-{b_2}^2-{d_1}^2+{a_1}^2+{b_1}^2$\\
The eqation (\ref{eq6}) can be reduced to
\begin{align}
x = \dfrac{z_0+2(b_2-b_1)y}{2(a_1-a_2)}\label{eq7}
\end{align}
For simplification, this can be written as
\begin{align}
x = z + py \label{eq8}
\end{align}

where $z=\dfrac{z_0}{2(a_1-a_2)}$ , and $p =\dfrac{2(b_2-b_1)}{2(a_1-a_2)}$ \\
Substituting the value of $x$ in equation (\ref{eq1}), we obtained
\begin{align}
 (p^2+1)y^2 + (2zp-2a_1p-2b_1)y - ({d_1}^2-{a_1}^2-{b_1}^2-{z}^2+2a_1z) = 0 \label{eq9}
\end{align}
Solving the quadratic equation (\ref{eq9}), let the values obtained be $y_1$ and $y_2$. Let $x_1$ and $x_2$ be the values corresponding to $y_1$ and $y_2$ respectively. 
Therefore, the proposed algorithm estimates two positions  $\mathit{P_{1}}$\textit{$(x_1,y_1)$}, and  $\mathit{P_2}$\textit{$(x_2,y_2)$}.

In order to select the correct estimated position a \textit{dead reckoning} approach is used. In this approach, a localized node say $\mathit{k}$ uses the location,
\textit{$p_{prev}$}
at the checkpoint $\mathit{t_i}$  to estimate its location in the next checkpoint at $\mathit{t_{i+1}}$.
Let $\mathit{v}$ be the velocity and $\mathit{t}$ be the time interval between the two successive checkpoints. 
Distance $\mathit{d}$ traveled by the node $\mathit{k}$ between two successive checkpoints is calculated as $\mathit{d = v * t}$. 
Therefore, at the checkpoint $\mathit{t_{i+1}}$, an unknown node knows its position at checkpoint $\mathit{t_{i}}$ and the distance $\mathit{d}$ traveled  
between the two successive checkpoints. Also, the node has two anchor positions, i.e., $(a_1,b_1)$, $(a_2,b_2)$. Then, the node uses trilateration to calculate 
the position $\mathit{P}$\textit{$(\hat{x},\hat{y})$}. Next, the node computes the correction factor $\mathit{Cf}$ to select one of the two estimated 
positions $\mathit{P_{1}}$ and $\mathit{P_{2}}$. The correctness factor is computed as:
\begin{center} 
$  Cf_1 = \sqrt{(\hat{x}-x_1)^2+(\hat{y}-y_1)^2} $\\
$  Cf_2 = \sqrt{(\hat{x}-x_2)^2+(\hat{y}-y_2)^2} $
 \end{center}
where $Cf_1, Cf_2$ represents the distance of position $\mathit{P_1}$, and $\mathit{P_2}$ from the position $\mathit{P}$ estimated \textit{via}
trilateration. The correct position of the node is $\mathit{P_{1}}$\textit{$(x_1,y_1)$} if ${Cf_1} < {Cf_2}$ else the correct position is
$\mathit{P_{2}}$\textit{$(x_2,y_2)$}. This is because, calculated position $\mathit{P}$\textit{$(\hat{x},\hat{y})$} always deviate from the
 actual position by a small margin.
Once a node is localized, it broadcasts beacons until the end of the checkpoint. 

\begin{figure}[!ht]
 \centering
\includegraphics[width=5.6in]{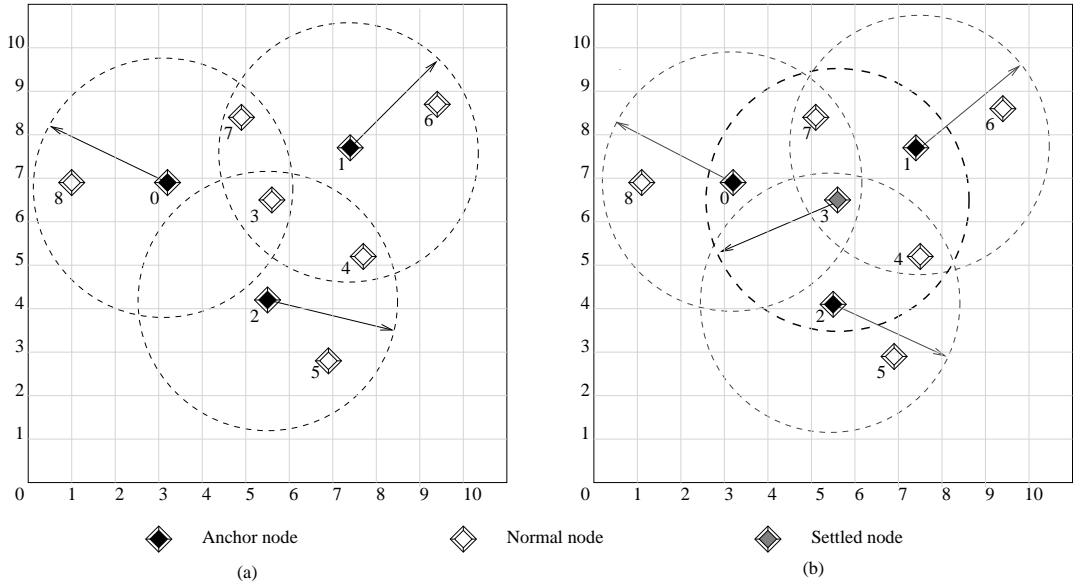}
\caption{ Initialization phase: (a) At the first checkpoint, anchor nodes transmit beacons,
 and normal nodes gets localized via trilateration; (b)  normal nodes that are short of 1 or 2 beacons gets localized with the help of settled nodes.} 
\label{fig:locproc}
\end{figure}

\begin{figure}[!h]
\centering
 \includegraphics[width=6.0in]{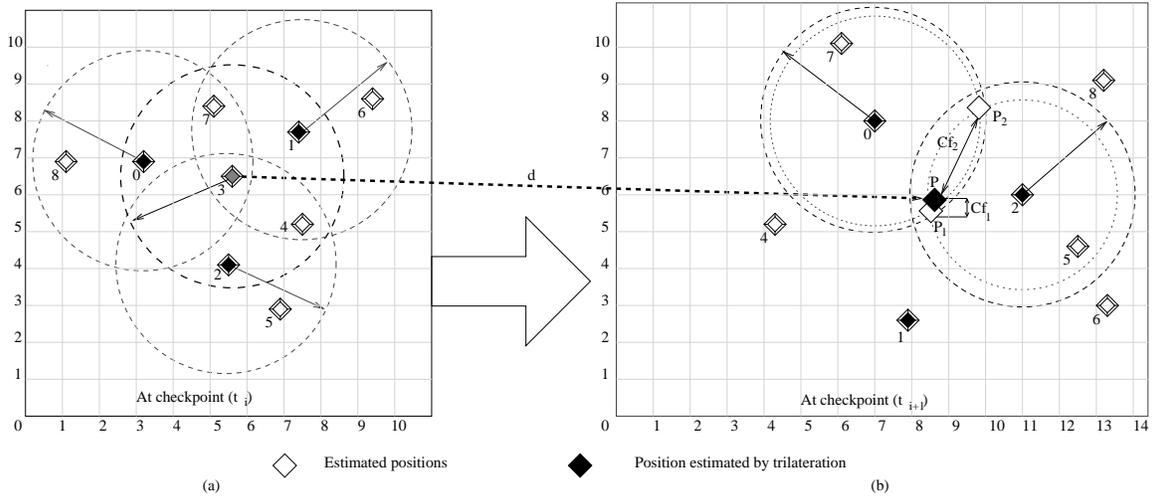}
\caption{Normal node 3 at checkpoint $\mathit{t_{i+1}}$ estimates two locations $\mathit{P_{1}}$ and $\mathit{P_{2}}$ using two anchor nodes 0 and 2. The
 correct position is selected  using the previous position of node 3 at checkpoint $\mathit{t_{i}}$.}
 \label{fig:deadreckon}
\end{figure}

We illustrate the localization process in proposed scheme using Fig. \ref{fig:locproc}. Localization in the initialization phase is shown in Fig. \ref{fig:locproc}(a) and 
\ref{fig:locproc}(b).
Node 3 in Fig. \ref{fig:locproc}(a) receives beacon from three anchor nodes $0, 1,$ and $2$ at checkpoint \textit{$t_{i}$} and gets localized. Nodes 4 and 7 receives
 only two beacons, whereas nodes 5, 6, and 8  receives
only one beacon. These nodes at this point of checkpoint  \textit{$t_{i}$} can not localize, as the number of beacons required for localization for the first time is \textit{three}.
Node 3 broadcast a beacon after localization. Nodes 4 and 7 gets localized after receiving beacon from node 3. This is shown in Fig. \ref{fig:locproc}(b).   
 This co-operative, distributive process of localization continues until the end of the checkpoint. At the end of the checkpoint \textit{$t_{i}$}, nodes 6 and
 8 have only one beacon. Both the nodes delete the received beacons and continues moving.

Fig. \ref{fig:deadreckon}(b) illustrate the sequent phase at checkpoint \textit{$t_{i+1}$}. 
We consider node 3, to explain localization using two anchor nodes. Let the co-ordinate of node 3 at checkpoint \textit{$t_{i}$} be $(5.5,6.3)$ as shown in
Fig. \ref{fig:deadreckon}(a), and the distance traversed during the checkpoint interval \textit{$t_{i}$} and \textit{$t_{i+1}$} be $3.5$ unit. At checkpoint 
\textit{$t_{i+1}$}, node 3 can be localized using two anchor nodes 0 and 2, as shown in Fig. \ref{fig:deadreckon}(b). Let the co-ordinates of node 0 and 2 be
$(7,8)$ and $(11,6)$ respectively as shown in Fig. \ref{fig:deadreckon}(b). Using B\'ezout's theorem node 3 estimates two locations $\mathit{P_{1}}(8.54,5.54)$ 
and $\mathit{P_{2}}(9.98,8.24)$. To select one of the above two locations dead reckoning approach is used. Based on the location of node 0, node 2 and its previous 
location, node 3 estimates its new location $\mathit{P}$\textit{$(\hat{x},\hat{y})$} equal to $(8.78,6.03)$ using trilateration technique. Then, node 3
calculates the correctness factor ${Cf_1}$ and  ${Cf_2}$ to find the least deviated estimated position from $\mathit{P}$\textit{$(\hat{x},\hat{y})$}.
The computed value of ${Cf_1}$ and ${Cf_2}$ is 0.738 and 2.514 respectively. Since ${Cf_1} < {Cf_2}$ the position $\mathit{P_{1}}(8.54,5.54)$ is selected as the 
correct estimated  position. It can be observed from the Fig. \ref{fig:deadreckon}(b) the actual position of node 3 is very close to the estimated position. 
 
The proposed localization algorithm is given as Algorithm 1. \\
\SetAlFnt{\scriptsize}
 \begin{algorithm}[H]
\DontPrintSemicolon
  \DontPrintSemicolon 
{\bf Notation:} $\mathcal{A}$: Anchor node, $\mathcal{U}$: Unknown node, $\mathcal{S}$: Settled node\;
{\bf Variables for $\mathcal{U}$ and  $\mathcal{S}$:}\;
 \lnlset{}{1}$beaconSet\gets \phi$\tcc*[r]{Set of received locations (x,y).}
 \lnlset{}{2}$locfirst\gets0$ \tcc*[r]{Set to 1, if $\mathcal{U}$ has completed initialization phase.}
 \lnlset{}{3}${P_{prev}}\gets -1$ \tcc*[r]{Stores current estimated position of $\mathcal{S}$ used in next checkpoint.}
 \lnlset{}{4}${Status}$ \tcc*[r]{Indicates node type: Its value can be $\mathcal{A}$ or $\mathcal{U}$ or $\mathcal{S}$ }
 /*{ \tt This algorithm is distributed and event oriented in nature. So, in each node type ($\mathcal{A}$, $\mathcal{U}$, $\mathcal{S}$) different actions get fired
 according to event triggered (recursive time interval for checkpoint or beacon reception).}*/\;
\BlankLine
  \textbf{\underline{For Anchor Node:}}\;
 \lnlset{}{5} \If{ $Status = \mathcal{A}$}{
 \lnlset{}{6} broadcast $beacon$\;
 \lnlset{}{7} exit \;
  }
\BlankLine
  \textbf{\underline{For Unknown/Settled Node:}}\;
 \lnlset{}{8}\eIf{ $Status = \mathcal{U}$}{
 \lnlset{}{9} Repeat the steps 10 and 11 until \;
 \lnlset{}{}   (!( $(sizeof(beaconSet) \geq 3)$  AND  $(locfirst = 0)$) OR !($(sizeof(beaconSet) \geq 2)$  AND  $(locfirst = 1)$)) AND !(End of Checkpoint)\;
 \lnlset{}{10}received  $beacon$\tcc*[r]{A node monitors the beacon until the end of checkpoint or either one of the following happens: 1. Unknown node
 has received three beacons, 2. Settled node has received two beacons.}
 \lnlset{}{11}$beaconSet\gets beaconSet\cup{\{beacon\}}$\;
 \lnlset{}{12}\If(\tcc*[f]{Initialization phase.}){ $(sizeof(beaconSet) \geq 3)$  and  $(locfirst = 0)$}{
\lnlset{}{13} $Position \gets Trilateration(beaconSet)$ \tcc*[r]{Localize using Trilateration.}
 \lnlset{}{14}broadcast $beacon$ \tcc*[r]{Broadcast estimated position.}
 \lnlset{}{15}${P_{prev}}\gets Position$\;
 \lnlset{}{16}$locfirst\gets1$\;
 \lnlset{}{17}$Status\gets \mathcal{S}$}
 
 \lnlset{}{18}\uElseIf(\tcc*[f]{Sequent phase.}){ $(sizeof(beaconSet) \geq 2)$  and  $(locfirst = 1)$} {
\lnlset{}{19}$Position\gets$ Use $beaconSet$ and ${P_{prev}}$\tcc*[r]{Localize using dead reckoning.}
\lnlset{}{20}broadcast $beacon$ \tcc*[r]{Broadcast estimated position.}
\lnlset{}{21}${P_{prev}}\gets Position$\;
\lnlset{}{22}$Status\gets \mathcal{S}$\;
 }
 }{
\lnlset{}{23}delete $beacon$  \tcc*[r]{$\mathcal{A}$ or $\mathcal{S}$ do not need beacon as these are localized already}
}
/*{ \tt Within different checkpoints a node remember only the values of ${P_{prev}}$ and $locfirst$.}*/\;
\lnlset{}{24}\While{((End of checkpoint) AND ($Status = \mathcal{S}$ OR $\mathcal{U}$))}{
\lnlset{}{25}$beaconSet\gets\phi$ \tcc*[r]{At the end of checkpoint delete received beacons.}
\lnlset{}{26}$Status\gets \mathcal{U}$ \tcc*[r]{For localizing in next checkpoint settled node changes status to $\mathcal{U}$.}}
\caption{\sc DRLMSN Localization algorithm}
\label{algo2}
\end{algorithm}
\textbf{Description of Algorithm \ref{algo2}:} \textit{The above algorithm is called at each node in the start of each checkpoint. 
In each checkpoint, anchor nodes broadcast beacons. This is mentioned in lines 5 -- 7 of Algorithm \ref{algo2}.
 Line 12 -- 17 localizes a node in the initialization phase. In this phase a node needs three beacons and gets localized via trilateration. Sequent phase localization is mentioned in
 lines 18 -- 22. In this phase a node require only two beacon node for localization. This phase uses dead reckoning approach for the unambiguous localization of unknown node.
 Lines 6, 14, and 20 in the algorithm does the job of beacon broadcast.
}
\section{Performance Evaluation}\label{eval}
\noindent
We have simulated the proposed scheme using Castalia simulator \cite{castalia} that runs on the top of
OMNET++. We have made the following assumptions in our simulation: \textit{(i)} nodes are considered to be homogeneous, with respect to transceiver power and receiver
sensitivity. This helps in controlling the connectivity between nodes in the network easily; \textit{(ii)} for simplicity,
 we consider transmission range of all the nodes as a perfect circle; and {\it (iii)} all the three beacon nodes are synchronized. This results in the accurate localization of unknown nodes which otherwise tries to localize
with obsolete beacons.

The key metrics used for evaluating the  localization algorithm is the accuracy in location estimation. We have calculated the estimated error 
as the difference between the estimated position and the actual position. The average root mean square error (RMSE) is 
calculated as:
\begin{center}$Average\hskip0.3em RMSE =\dfrac{\sum_{i=1}^{N-P}||\hat\theta_i - \theta_i ||} {N-P}$\end{center} 
where $\hat\theta_i $ is estimated position, $\theta_i$ is actual position, $N$ is the total number of nodes in the network, 
and $P$ is number of anchor nodes.

We have considered the following parameters in our simulation: (\textit{i}) nodes are randomly deployed in a sensor field of area $200 \times 200$ $m^2$; (\textit{ii})
symmetric communication within a communication range of 20 meters; (\textit{iii}) anchor node density is 10\%. We have defined the anchor density as the ratio
between he anchor nodes to the total nodes in the network; (\textit{iv}) transmission power is -5 dBm; (\textit{v}) path loss exponent ($\eta$) is
2.4; and (\textit{vi}) modified random waypoint mobility model \cite{rwp} and random direction mobility model \cite{directionmodel}. 
We have compared DBNLE with another range based localization technique called RSS-MCL  proposed by Wang and Zhu \cite{comp}.
Through simulation, we studied the impact of mobility model, anchor density, node speed, number of normal nodes, and deployment topology on location estimation.
Each of the above parameter is explained below.
 
 \begin{figure}
\begin{minipage}{0.47\textwidth}
   \includegraphics[width=3.3in]{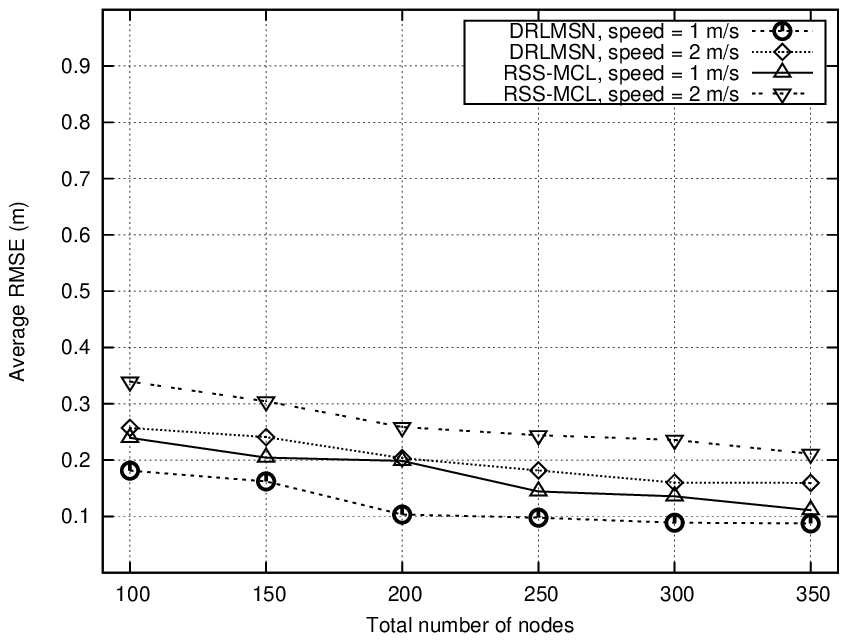}
\caption{Impact of increase in the number of nodes on the localization error in modified random waypoint mobility model.} 
\label{fig:nodesrw}
 \end{minipage}
\hspace{0.5cm}
\begin{minipage}{0.47\textwidth}
 \includegraphics[width=3.3in]{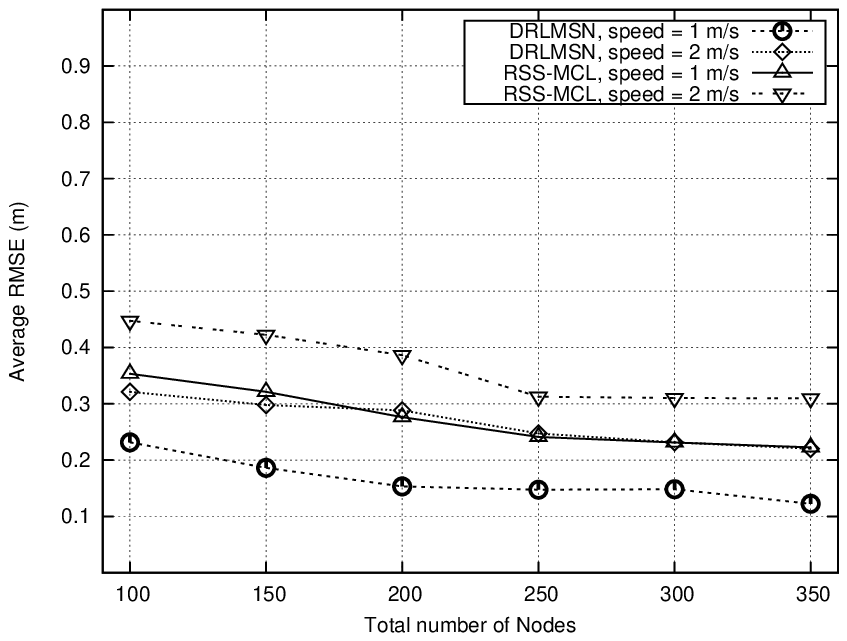}
\caption{Impact of increase in the number of nodes on the localization error in random direction mobility model.} 
\label{fig:nodesrd}
 \end{minipage}
\end{figure}
\begin{figure}
\begin{minipage}{0.47\textwidth}
   \includegraphics[width=3.2in]{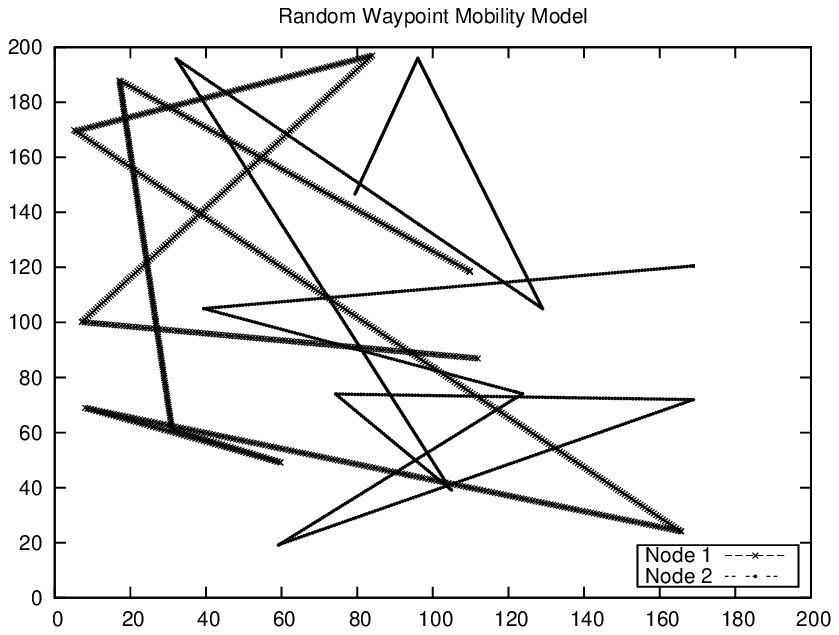}
\caption{Mobility pattern of nodes in random waypoint mobility model.} \label{fig:waypointmodel}
 \end{minipage}
\hspace{0.5cm}
\begin{minipage}{0.47\textwidth}
\centering{
 \includegraphics[width=3.2in]{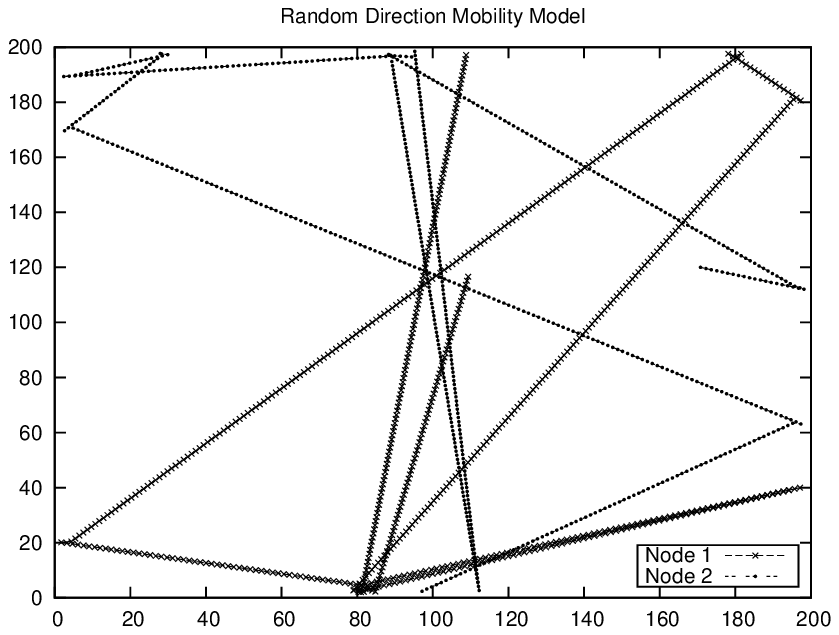}
\caption{Mobility pattern of nodes in random direction mobility model.} \label{fig:directionmodel}}
 \end{minipage}
\end{figure}
\textbf{Impact of Mobility Model:}
Mobility pattern plays an important role in the localization process. Besides increasing the network connectivity and coverage area, mobility affects 
the localization accuracy, also drains the battery quickly, and the percentage of localized nodes. We considered two mobility models
(\textit{i}) Random Waypoint Mobility Model (RWMM), (\textit{ii}) Random Direction Mobility Model (RDMM) and have shown the effect
of mobility model on localization accuracy.

In RWMM, a node  randomly chooses a new destination in a direction between [0, 2$\pi$] and moves towards that destination  with a speed in 
the range [$v_{min}, v_{max}$]. While in RDMM, a node randomly chooses a direction between [0, 2$\pi$], a speed in the range [$v_{min}, v_{max}$] 
and moves in the chosen direction upto the boundary of the network. After reaching the boundary same process is repeated. In both RWMM and RDMM, a
node pauses for some predefined time before changing its direction. We have set the pausetime to be zero, in order to simulate a continuous mobility model.
From Fig. \ref{fig:nodesrw}  it is observed that RWMM have lower average RMSE than RDMM 
 as shown in Fig. \ref{fig:nodesrd}. The reason for this difference in error is due to mobility pattern of nodes. In RWMM, nodes 
mostly move within the vicinity of the center. They are less likely to move towards the boundaries of the network as shown in Fig. \ref{fig:waypointmodel}.
Therefore, a node will have relatively higher number of neighbours. As a result, a normal node selects the most nearest neighbours
which results in lesser inaccuracy. In contrast to this, a node moves uniformly throughout the field in RDMM as shown in Fig. \ref{fig:directionmodel}. This
type of movement does not favor the selection of best neighbours, because a node is surrounded by lesser number of neighbours. It is observed
 from the Fig. \ref{fig:nodesrw} and \ref{fig:nodesrd} that average
 RMSE is lesser in DBNLE compared to RSS-MCL. This is because in RSS-MCL the number of beacons used to filter the generated sample is more compared to DBNLE which 
 uses only 2--3 beacons. Due to mobility, increasing dependency on the number of beacons used increases the uncertainty in position estimation.  
Among the two mobility models, RDMM increases network coverage while RWMM increases the connectivity among nodes.

\begin{figure}
\begin{minipage}{0.47\textwidth}
   \includegraphics[width=3.1in]{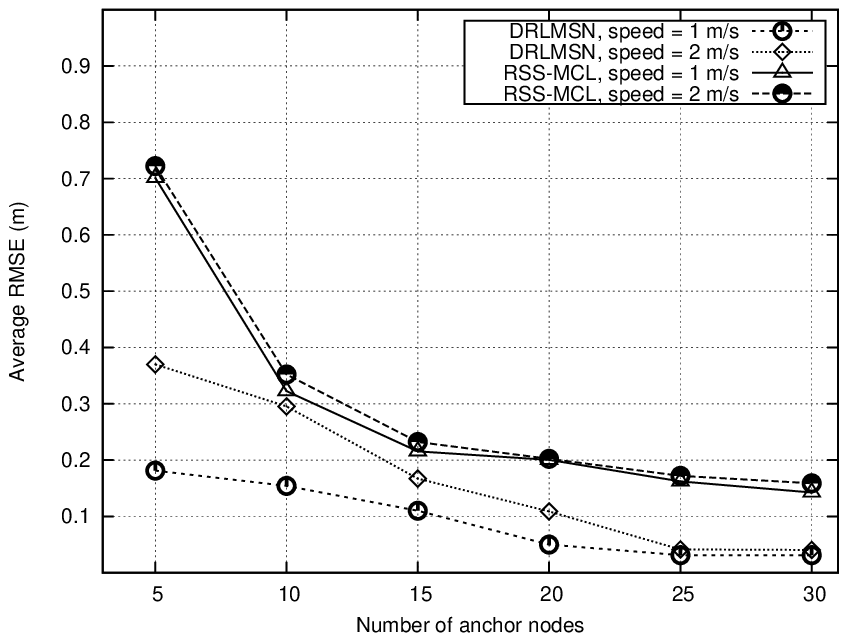}
\caption{Impact of increase in anchors on the localization error in modified random waypoint mobility model.} 
\label{fig:anchorrw}
 \end{minipage}
\hspace{0.5cm}
\begin{minipage}{0.47\textwidth}
\includegraphics[width=3.1in]{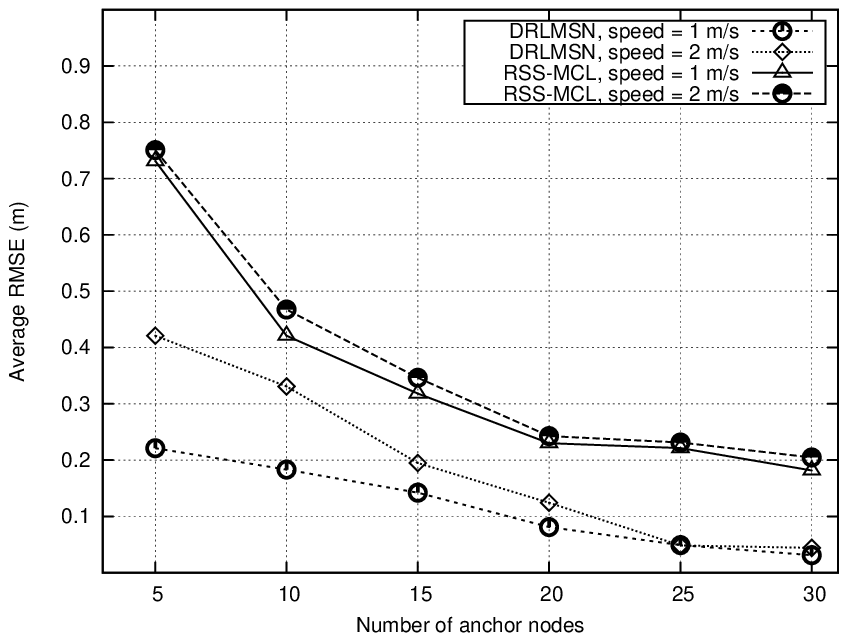}
\caption{Impact of increase in anchors on the localization error in random direction mobility model.} 
\label{fig:anchorrd}
 \end{minipage}
\end{figure}
\textbf{Impact of anchor nodes:}
For a fixed network size, increasing the anchor density results in the localization of more nodes in lesser time. This is because, most of the nodes
obtain higher number of anchor nodes as their neighbour. To find the effect of anchor density on localization error we varied the anchor density
between 5\% to 20\% keeping the total number of nodes to be fixed at 200. The plot for anchor density \textit{vs.} localization error in RWMM and RDMM is shown in Fig. \ref{fig:anchorrw} and 
\ref{fig:anchorrd} respectively. It is observed from the figures that the average RMSE  decreases with the increase in the anchor density. This is because:
(\textit{i}) higher the anchor density, lesser the number of nodes to be localized; (\textit{ii})  a node gets more number of accurate beacons - resulting 
in lesser error accumulation and propagation. It is also observed that with an increase in anchor density the average rate of decrease of RMSE is higher, 
and at a higher anchor density the rate of decrease is lesser. Increase in the number of anchors do not affect average RMSE to a greater extent in RSS-MCL as compared to DBNLE.
This is because in RSS-MCL position estimation depends heavily on the quality of sample generation where as in DBNLE it directly depends on the number of beacons received.
Furthermore, average RMSE is lesser in RWMM as compared RWDM. This is attributed to the neighbor density.
 In RWMM, a node has higher neighbour density as compared to RDMM.

\textbf{Impact of node speed:}
The effect of speed on average RMSE by varying the anchor and node density is shown in Fig. \ref{fig:nodesrw}, \ref{fig:nodesrd},
\ref{fig:anchorrw}, and \ref{fig:anchorrd}.
It is observed from the above figures that with increase in speed the localization error also increases.
Above figures show that the location estimation of a node in mobile WSNs is greatly affected by the node speed.
 A node covers more distance per unit time at higher speed. This increase in speed results in: \textit{(i)} 
increase in the uncertainty of localizing a node accurately, as the area over which a node needs to be localized increases, \textit{(ii)} 
with the increase in distance covered, multi-path fading and shadowing comes into play. This affects the distance measurements
and decreases the efficiency of range based localization algorithm, \textit{(iii)} it also affects the basic functionality, \textit{i.e.}, sensing is not
 properly done when a node moves too fast, \textit{(iv)} it increases the localization percentage in low anchor
density networks because increase in speed increases the network coverage.
 
 \begin{figure}
\begin{minipage}{0.47\textwidth}
   \includegraphics[width=3.2in]{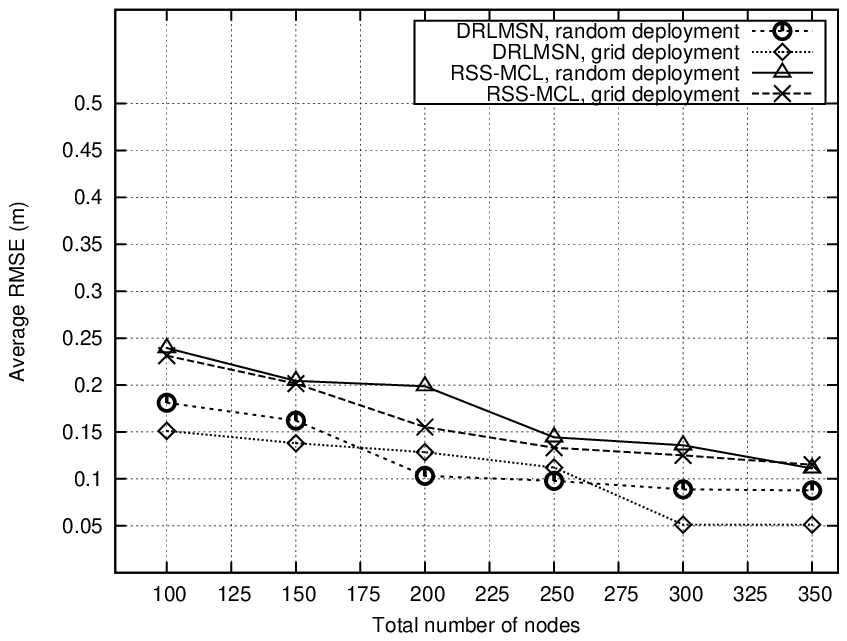}
 \caption{Impact of node deployment on the localization error in modified random waypoint mobility model.} 
\label{fig:gridrw}
 \end{minipage}
\hspace{0.5cm}
\begin{minipage}{0.47\textwidth}
\includegraphics[width=3.2in]{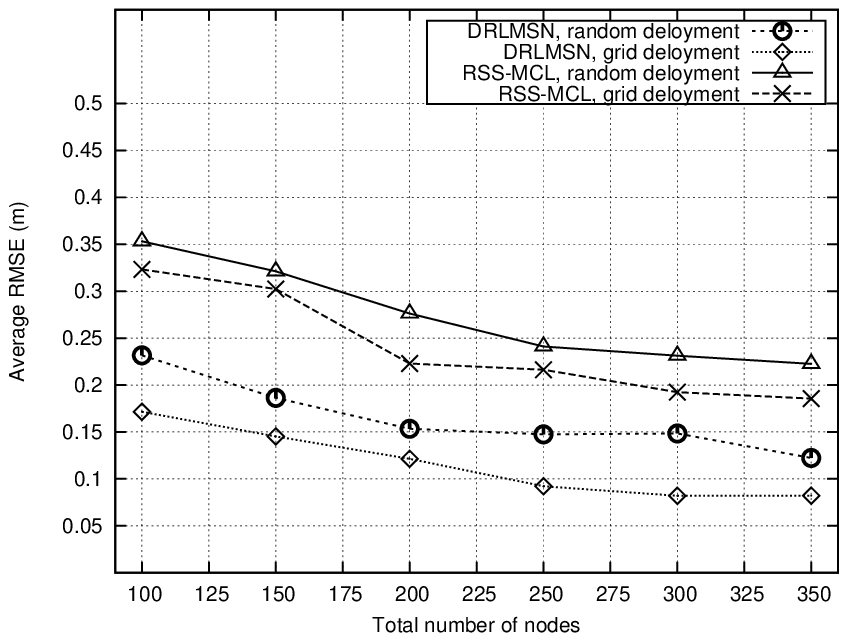}
 \caption{Impact of node deployment on the localization error in random direction mobility model.} 
\label{fig:gridrd}
 \end{minipage}
\end{figure}
\textbf{Impact of normal nodes:}
The plot for normal nodes \textit{vs.} localization error is shown in Fig. \ref{fig:nodesrw} and \ref{fig:nodesrd} respectively.
With increase in the number of normal nodes there is a significant increase in the percentage of localized nodes. This also
results in the decrease of localization time and localization error. Decrease in localization time is attributed to more number of localized neighbours of 
a normal node. It is observed from the Fig. \ref{fig:nodesrw} and \ref{fig:nodesrd} that localization error decreases gradually with the
increase of nodes. The reason for this decrease is the selection of more number of nearest in-range neighbours. Closer is the neighbour lesser is the
ranging error; as the quality of signal (RSSI) is directly affected by the distance between the transmitter and receiver node.

\textbf{Impact of deployment/topology of nodes:} 
Next, we consider the effect of deployment on localization error. We have considered two deployment scenarios: (\textit{i}) random, and (\textit{ii}) grid
to study their effect on localization error.
It is observed that in some cases nodes do not localize early and takes 
longer time to localize. Consequently, this increases the localization time of whole network.
This is due to the non-availability of requisite number of beacons for localization.
 One of the major reason for this is the way in which nodes
are deployed initially and the manner in which nodes move.
 It is observed that if nodes are randomly deployed, then 30\% of the nodes fail to localize
in the first 2 to 3 checkpoints, whereas in grid network around 90\% of nodes localize
in the first checkpoint itself. In the next checkpoint all nodes get localized. From the Fig. \ref{fig:gridrw} and \ref{fig:gridrd}, 
it is observed that the localization error is lesser in grid deployment than in random deployment.

\begin{figure}
\begin{minipage}{0.47\textwidth}
\includegraphics[width=3.1in]{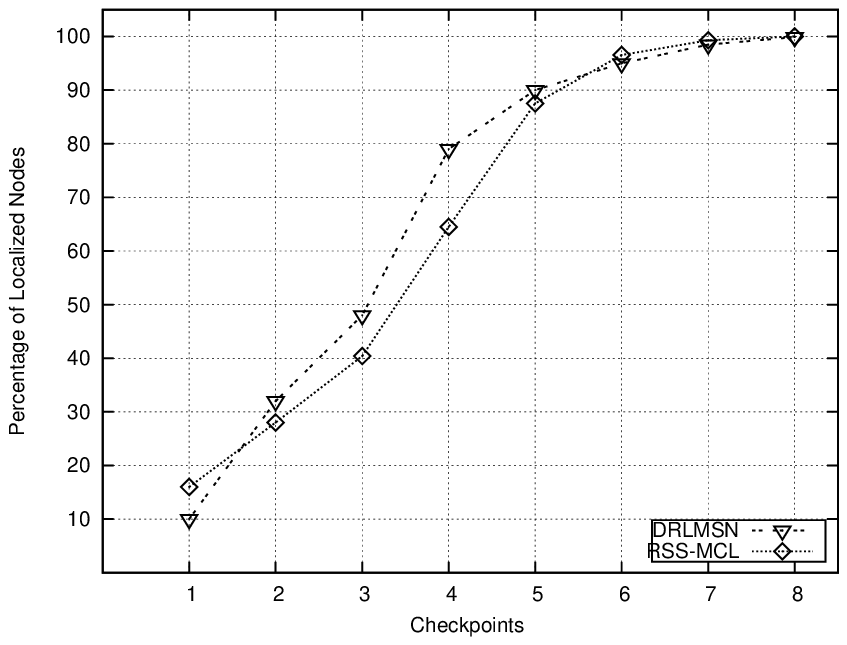}
\caption{Percenatge of localized nodes at successive checkpoints in modified random waypoint mobility model.}
 \label{fig:percentrw}
 \end{minipage}
\hspace{0.5cm}
\begin{minipage}{0.47\textwidth}
 \includegraphics[width=3.1in]{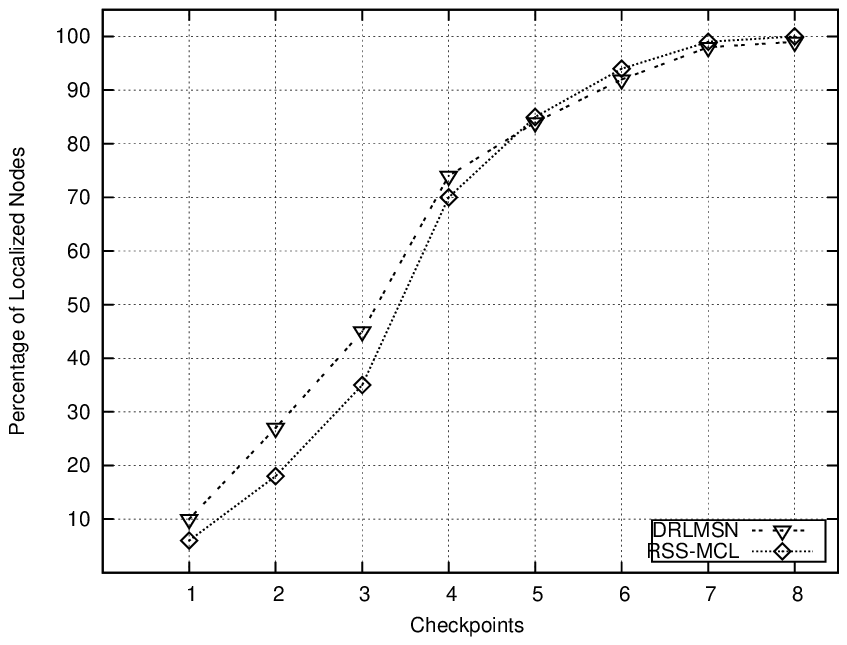}
 \caption{Percenatge of localized nodes at successive checkpoints in random direction  mobility model.}
  \label{fig:percentrd}
 \end{minipage}
\end{figure}
Finally, we studied the percentage of nodes localized at different checkpoints. The plot for percentage of localized nodes \textit{vs.} checkpoints is shown in
Fig. \ref{fig:percentrw} and \ref{fig:percentrd}. It is observed that the percentage of nodes localized increases as the checkpoint increases. Majority 
of the nodes gets localized after the fourth checkpoint. Percentage of localized nodes in RSS-MCL is relatively lesser as compared in DRLMSN. 
This is because the
time spent in  sample generation and filtering is more than checkpoint duration. As a result, most of the nodes fail to localize in RSS-MCL due to this time constraint.

\section{Conclusion}\label{con}
\noindent
 A large number of localization techniques have been developed for static WSNs. These techniques can not be applied to mobile WSNs. Only a few 
localization techniques has been proposed for mobile WSNs. Most of these techniques considered either normal node or anchor nodes to be static.
In this paper we propose a technique called dead reckoning localization for mobile WSN. We have considered both the normal nodes and anchor nodes to be mobile.
 As the nodes move in a sensor field, their position changes with time. Therefore, a mobile node has to be localized as long as it is alive.
In the proposed technique, nodes are localized at discrete time intervals called checkpoints. A normal node is localized for the first time using
three anchor nodes. For their subsequent localizations only two anchor nodes are used and a dead reckoning technique is applied. Reduction in the number of anchor
nodes required for localization from three to two result in faster localization  and lesser localization error. We have compared the proposed scheme with an existing 
similar scheme called RSS-MCL. It is observed that the proposed scheme has faster localization time with lesser localization error than RSS-MCL. 
We have also studied the impact of node density, anchor density, node speed, deployment type and mobility pattern on localization. It was observed that the above parameters
have strong influence in the localization time and localization error.
  
In future we would like to check the validty of DRLMSN with different mobility models other than RWMM and RDMM. Also, to implement the proposed
 scheme in an real environment and check its performance.
 \bibliographystyle{unsrt}	
  \bibliography{recokon}


\end{document}